\documentclass[aps,prl,longbibliography,twocolumn,reprint,amsmath,amsfonts,superscriptaddress]{revtex4-1}

\usepackage{graphicx}

\newcommand{\eqnref}[1]{Eq.\ (\ref{#1})}
\usepackage{amsmath}
\usepackage{hyperref}
\usepackage{color}
\usepackage{framed}

\newcommand{\bra}[1]{\langle #1|}
\newcommand{\ket}[1]{|#1\rangle}
\newcommand{\braket}[2]{\langle #1|#2\rangle}

\usepackage{amsthm}
\newtheorem{lemma}{Lemma}
\theoremstyle{definition}
\newtheorem{thm}{Theorem}

\newcommand{\oned}{1\nobreakdash-D}
\newcommand{\twod}{2\nobreakdash-D}

\begin{document}

\title{Symmetry-protected phases for measurement-based quantum computation}

\author{Dominic V. Else}
\affiliation{Centre for Engineered Quantum Systems, School of Physics,
The University of Sydney, Sydney, NSW 2006, Australia}
\author{Ilai Schwarz}
\affiliation{Centre for Engineered Quantum Systems, School of Physics,
The University of Sydney, Sydney, NSW 2006, Australia}
\affiliation{Racah Institute of Physics, The Hebrew University, Jerusalem 91904,
Israel}
\author{Stephen D. Bartlett}
\affiliation{Centre for Engineered Quantum Systems, School of Physics,
The University of Sydney, Sydney, NSW 2006, Australia}
\author{Andrew C. Doherty}
\affiliation{Centre for Engineered Quantum Systems, School of Physics,
The University of Sydney, Sydney, NSW 2006, Australia}

\begin{abstract}
  Ground states of spin lattices can serve as a resource for measurement-based
  quantum computation. Ideally, the ability to perform quantum gates via
  measurements on such states would be insensitive to small variations in the
  Hamiltonian. Here, we describe a class of symmetry-protected
  topological orders in one-dimensional systems, any one of which ensures the perfect operation of the identity
  gate. As a result, measurement-based quantum gates can be a robust property of an entire phase in a quantum spin lattice, when protected by an appropriate symmetry. 
\end{abstract}

\maketitle

Quantum computation exploits quantum entanglement to achieve computational
speedups. However, creating entanglement between particles in a sufficiently
controlled way to allow for quantum computation has proved a major technical
challenge. One potential approach is \emph{measurement-based quantum
computation} (MBQC) \cite{raussendorf-prl-2001, *raussendorf_et_al_2003,
briegel_mbqc_natphys}, where
universal quantum computation is achieved by means of non-entangling operations
(namely, single-particle measurements) on an already entangled resource state.
The resource state need not be prepared
coherently; instead, one could imagine constructing interactions between
neighboring spins on a lattice, governed by a gapped Hamiltonian whose ground
state is a universal resource state for MBQC
\cite{doherty-bartlett-prl-2008,chen_tricluster,miyake_aklt,*raussendorf_aklt1}. For this approach to
be robust, the capability of ground states to serve as a resource for MBQC
would have be insensitive to small variations in the Hamiltonian, like a form
of quantum order \cite{doherty-bartlett-prl-2008}.

In this Letter, we draw an explicit connection between MBQC and a type of
quantum order called \emph{symmetry-protected topological order} (SPTO)
\cite{gu_wen_2009,chen_gu_wen,schuch}.
Specifically, we will describe a class of quantum phases in which the perfect
operation of the identity gate in MBQC can be derived directly from the
presence of SPTO; consequently, this perfect operation
is a robust property which is maintained throughout the entire phase.
Our results will be expressed in the context of one-dimensional systems.  Such
systems are not expected to allow for universal MBQC, but the ground states of
certain 1\nobreakdash-D spin chains can be used as \emph{quantum computational wires}~\cite{gross-eisert-2010-webs}, meaning, loosely, that through
single-particle
measurements one can propagate a logical qubit down the chain while applying
single-qubit unitaries. Later, we will also explain how our results can
be applied to higher-dimensional systems (which can allow for universal MBQC) by considering them as `quasi-1D'.

A well-known example of a one-dimensional system whose ground state can serve as
a quantum computational wire is the Affleck-Kennedy-Lieb-Tasaki (AKLT) antiferromagnetic
spin-1 chain~\cite{aklt2,*aklt1,brennen-aklt-mbqc-2008}. This system lies in a quantum
phase, called the \emph{Haldane phase}, characterized by SPTO and protected by a
$Z_2 \times Z_2$ rotation symmetry~\cite{pollmann-arxiv-2009,pollmann-prb-2010}, so that no symmetry-respecting path of local Hamiltonians
can interpolate between the Haldane phase and a product state without crossing a
phase transition.
The perfect operation of the identity gate throughout
the Haldane phase has been noticed before in various guises
\cite{uq_teleportation,bartlett_renormalization}, as well as the strictly weaker
condition of diverging localizable entanglement length
\cite{venuti-2005-le-string}. It should be emphasized that in MBQC, repeated application of the identity gate corresponds to the propagation of a logical state arbitrarily far down the chain without error. Thus, the identity gate is not a null operation in this context, and its perfect operation is a striking and non-trivial property of the Haldane phase.

Our purpose in this Letter will be to show explicitly how the perfect operation
of the identity gate
arises as a direct manifestation of SPTO. As a result, we can
apply our technique more generally to a whole class of quantum phases
characterized by SPTO, including phases containing the \oned{} cluster state;
qudit cluster states \cite{qudit_cluster_state}; and cluster states in higher
dimensions.  In addition, we show that gates other than the identity are not
expected to exhibit similar robustness, explaining the numerical observations in
Ref.~\cite{bartlett_renormalization}.  

\emph{Symmetry-protection of the identity gate in correlation space.---}
The connection between SPTO
and MBQC will be expressed through the correlation space picture of
\cite{correlation_space_pra}, which is a
particularly natural way to understand the operation of gates in \oned{} resource states.
This picture assumes a resource state $\ket{\Psi}$ that can be represented as a \emph{matrix-product state} (MPS), 
\begin{multline}
  \label{mps}
  \ket{\Psi} = \sum_{k_1,\ldots,k_N} \bra{R} A[k_N] A[k_{N-1}] \cdots
  A[k_1] \ket{L} \\
  \times \ket{k_1,\ldots,k_N},
\end{multline}
where each $A[k_j]$, $k_j = 1,\ldots,d$ is a linear operator acting on a $D$-dimensional vector
space (known as the \emph{correlation space}), $\ket{L}$ and $\bra{R}$ are
states in correlation space, and $d$ is the dimension of the Hilbert space of
each spin. Here we are assuming translational invariance, for notational
simplicity only. When a projective measurement is performed on the first spin, with outcome $\ket{\psi}$, the effect
is to remove the first spin from the chain and induce an
evolution $\ket{L} \to A[\psi] \ket{L}$ on the correlation system,
where we use the notation
$A[\psi] = \sum_k A[k] \braket{\psi}{k}$.

As an introduction to our result, we will first state it for the special case of
the Haldane phase. One system within this phase is the spin-1 AKLT chain, for
which the ground state has an exact MPS representation of the
form~(\ref{mps}), with $D = 2$.
Expressed in the basis $\{ \ket{x}, \ket{y}, \ket{z} \}$, where
$\ket{\alpha}$ is the zero eigenstate of the spin-1 operator $S_\alpha$ for $\alpha = x,y,z$,
we have $A^{\rm AKLT}[\alpha] = \sigma_\alpha$, where $\sigma_{\alpha}$ are the Pauli spin operators.
Thus, the AKLT state has the particular property that there exists a basis, namely the
$\{ \ket{x}, \ket{y}, \ket{z} \}$ basis, such that measurements in this basis
induce an identity evolution (up to Pauli by-products) on the correlation
system.
Additionally, by measuring in a basis corresponding to a rotated set of axes,
it is possible to execute any single-qubit rotation in
correlation space (up to Pauli by-products) \cite{brennen-aklt-mbqc-2008}.
Therefore, the AKLT state can be said to act as a quantum computational wire.

We will now extend our correlation-space analysis beyond the AKLT chain to other
ground states within the Haldane
phase. We confine our discussion to states that can be exactly represented as
an MPS with a bond dimension $D$ that is independent of the system size.
Because arbitrary gapped ground states can be approximated by MPS
\cite{mps_faithfully,*hastings_area_law}, we expect that our discussion will
apply also to arbitrary systems in the Haldane phase.

The Haldane phase containing the AKLT chain is protected by the $Z_2
\times Z_2$ symmetry generated by the $\pi$ rotations about three orthogonal
axes.
The action of this symmetry on a spin-1 chain can be written as a
tensor product $[u(g)]^{\otimes N}$, where $N$ is the number of spins, and
$u(g)$ is the appropriate single-spin rotation operator for each group element
$g$ in the symmetry group $G = Z_2 \times Z_2$. We therefore refer to it as an
on-site symmetry.

In general, the invariance of a ground state under such an
on-site symmetry
 leads to symmetry constraints on the MPS tensor
$A[\cdot]$ used to construct the state's MPS representation
\cite{string_order_symmetries,chen_gu_wen,schuch,tensor_network_global_symmetry}; we will exploit
these constraints to prove our result.
Specifically, under an injectivity assumption which we expect to be satisfied in
a gapped phase, we have \cite{string_order_symmetries,schuch,chen_gu_wen}
\begin{equation}
\label{mps_symmetry_condition}
V(g)^{\dagger} A[\ket{\psi}] V(g) =
\beta(g) A[u(g)^{\dagger} \ket{\psi}],
\end{equation}
where $V(g)$ is some projective representation of $G$ acting on the correlation
system, and $\beta(g)$ is a one-dimensional linear representation of $G$. Now,
in general $V(g)$ can be decomposed as a tensor sum of irreducible projective
representations as $V(g) = \bigoplus_J V_J(g) \otimes \mathbb{I}_{m_J}$, where
$m_J$ is the multiplicity of the irrep $J$ in $V$.  For any ground state in the
Haldane phase, it is a consequence of Lemma \ref{uniqueness_lemma} below that
only one
irrep $\widetilde{V}(g)$ (of dimension 2) appears in this decomposition, so that 
\begin{equation}
  \label{group_representation_decomposition}
  V(g) = \widetilde{V}(g) \otimes \mathbb{I}_{\mathrm{junk}} . 
\end{equation}
That is, we have a tensor product decomposition of the
correlation system into a \emph{protected subsystem} [on which $V(g)$ acts
irreducibly as $\widetilde{V}(g)$] and a \emph{junk subsystem} (on which $V(g)$
acts trivially). The states $\ket{x}, \ket{y}, \ket{z}$ are
simultaneous eigenstates of all the elements $u(g)$. By an argument involving
Schur's Lemma (given in greater generality in Theorem \ref{mainthm}), it follows
that the tensor $A$ appearing in the MPS representation of the ground state must take the form
\begin{equation}
  \label{mps_tensor_decomposition_aklt}
  A[\alpha] = \sigma_\alpha \otimes A_{\mathrm{junk}}[\alpha], \quad \alpha = x,y,z,
\end{equation}
for some set of operators $A_{\mathrm{junk}}[\alpha]$ acting on the junk
subsystem. Recall that $A[\alpha]$ is the evolution induced on the correlation
system when a projective measurement results in the outcome $\ket{\alpha}$.
Thus, \eqnref{mps_tensor_decomposition_aklt} shows that
the ability to induce an identity evolution 
 in the protected subsystem (up to Pauli byproducts, dependent
on the measurement outcome but independent of the resource state) by measuring in the $\{ \ket{x}, \ket{y},
\ket{z} \}$ basis is dictated by the symmetry properties of the MPS tensor;
it is a property not just of the AKLT state, but rather of the entire Haldane phase.

Another state which can serve as a quantum computational wire is the \oned{} cluster
state, which is the ground state on a row of qubits of the local Hamiltonian $H = -\sum_i Z_{i-1} X_i Z_{i+1}$.
Like the AKLT state, the cluster state has an exact MPS representation, and it
lies within a symmetry-protected phase with respect to a $Z_2 \times Z_2$
symmetry \cite{son-cluster-2011}, in this case generated
by $\prod_{i \, \mathrm{even}} X_i$ and $\prod_{i \, \mathrm{odd}} X_i$.
We can
treat this symmetry as on-site provided that we group pairs of qubits into
sites.
The simultaneous eigenspace of the on-site symmetry representation is then $\{ |{+}{+}\rangle, |{+}{-}\rangle, |{-}{+}\rangle,|{-}{-}\rangle\}$, where $\ket{\pm} = \frac{1}{\sqrt{2}}(\ket{0} \pm \ket{1})$ (we
emphasize that this is a product basis, so that blocking sites does not
change the single-qubit nature of the measurements).  Identical to the AKLT case above, we again find that the ability to perform the identity gate by measuring in the appropriate basis is maintained
throughout the phase.
Similar results hold for the generalization of the
cluster state to $d$-dimensional particles \cite{qudit_cluster_state},
for which the relevant symmetry group is $Z_d \times Z_d$.

\emph{General statement of the result.---}We will now give the statement and
proof of our
result in a general setting. We consider a ground state that is
invariant under an on-site symmetry $[u(g)]^{\otimes N}$, where $u(g)$ is a
representation of some symmetry group $G$. We assume the ground state has an MPS representation satisfying the symmetry condition~(\ref{mps_symmetry_condition}), and we absorb $\beta(g)$ into $u(g)$
so that $\beta(g) = 1$.
A projective representation $V(g)$ is
characterized by its \emph{factor system} $\omega$,
such that
\begin{equation}
  V(g) V(h) = \omega(g,h) V(gh).
\end{equation}
An equivalence class of factor systems related by rephasing of the operators
$V(g)$
is called a \emph{cohomology class}, and we denote the cohomology class
containing a given factor system $\omega$ as $[\omega]$.
It was argued in Refs.\ \cite{chen_gu_wen,schuch} that each
cohomology class of $G$ corresponds to a distinct symmetry-protected phase.
For example, in the case of the MPS $A^{\rm AKLT}[\alpha] = \sigma_\alpha$ for the AKLT state,
where $G = Z_2 \times Z_2 = \{1,x,y,z\}$, it can be verified that
\eqnref{mps_symmetry_condition} is satisfied with the Pauli projective representation $V(1)=I$ and 
$V(\alpha) = \sigma_\alpha$ for $\alpha = x,y,z$. This corresponds to
a nontrivial cohomology class [not containing the trivial factor system
$\omega(g,h) = 1$], so that the AKLT chain lies in a
nontrivial symmetry-protected phase.

We now relate the symmetry condition (\ref{mps_symmetry_condition}),
which holds throughout the entire symmetry-protected phase, to the operation of
gates in the correlation-space picture.
We consider the case where the symmetry group $G$ is
a finite abelian group. For
simplicity, we will focus on the case where the cohomology class $[\omega]$ characterizing the
symmetry-protected phase is of a particular type. (An analogous result holds
for all non-trivial cohomology classes, but the structure of
correlation space is more involved in that case.) In particular, we consider the
case where the factor systems contained in $[\omega]$ are
\emph{maximally non-commutative}, meaning that the subgroup $G(\omega)= \{ g \in G |
\omega(g,h) = \omega(h,g)\ \forall\ h \in G\}$ is trivial. (Note, this
condition does
not depend on the choice of the representative $\omega$.) Under these
conditions, our main result can be
stated as follows:

\begin{framed}
\begin{thm}
\label{mainthm}
Consider a symmetry-protected phase characterized by a finite abelian symmetry
group and a maximally non-commutative
cohomology class $[\omega]$. Then for any MPS in this phase,
there exists a decomposition of the correlation system into protected and junk
subsystems, and a site basis $\{ \ket{i} \}$, such that measuring in the
basis $\{ \ket{i} \}$ leads to an identity gate evolution on the protected
subsystem 
up to an outcome-dependent byproduct $B_i$. That is to say, the MPS tensor $A$ has the
decomposition
\begin{equation}
\label{general_mps_decomposition}
A[i] = B_i \otimes A_{\mathrm{junk}}[i].
\end{equation}
The byproduct operators $B_i$ are unitary and are elements of a finite group.
Furthermore, they are the same for all possible MPS in the symmetry-protected phase.
\end{thm}
\end{framed}
For example, the factor system for the Pauli projective representation of $Z_2
\times Z_2$ is maximally non-commutative, and
\eqnref{mps_tensor_decomposition_aklt} is a special case of
\eqnref{general_mps_decomposition}.

\emph{Proof of Theorem \ref{mainthm}}.---
We will make use of the
following consequences of maximal non-commutativity of a factor system:
\begin{lemma}
Let $\omega$ be a maximally non-commutative factor system of a finite abelian
group $G$.
For every linear
character $\chi$ of $G$, there exists an element $h_\chi \in G$ such that, for
any projective representation $V(g)$ with factor system $\omega$,
\begin{equation}
  \label{h_chi_eqn}
  V(h_\chi) V(g) = \chi(g) V(g) V(h_\chi),
\end{equation}
\begin{proof}
We define a homomorphism $\varphi : G \to G^{*}$, where $G^{*}$ is the group of
linear characters of $G$, according to $[\varphi(h)](g) = \omega(h,g) \omega(g,h)^{-1}$. 
(That $\varphi(g) \in G^{*}$ for all $g$, and $\varphi$ is a homomorphism,
follows from the associativity condition satisfied by $\omega$, e.g.\ see Lemma
7.1 in \cite{kleppner_multipliers}).  Because the kernel of $\varphi$ is
$G(\omega)$, which is trivial by assumption, and $|G| = |G^{*}|$ for finite
abelian groups, it follows that $\varphi$ is invertible. We then set $h_\chi =
\varphi^{-1}(\chi)$. It can be checked that this satisfies \eqnref{h_chi_eqn}.
\end{proof}
\end{lemma}

\begin{lemma}
\label{uniqueness_lemma}
For each maximally non-commutative factor system $\omega$ of a finite abelian
group $G$, there exists a
unique (up to unitary equivalence) irreducible projective representation
$\widetilde{V}(g)$ with factor system $\omega$. The dimension of this irreducible representation is
$\sqrt{|G|}$.
\begin{proof}
See \cite{frucht1932,berkovich_characters}.
\end{proof}
\end{lemma}

For an MPS tensor $A$ satisfying the symmetry condition~(\ref{mps_symmetry_condition}), Lemma~\ref{uniqueness_lemma} implies that there
exists a tensor product decomposition of the correlation system into a protected and a junk subsystem such that $V(g)$ acts within the protected subsystem as
$\widetilde{V}(g)$ as in \eqnref{group_representation_decomposition}.  

Now we can prove Theorem \ref{mainthm}. We choose the measurement basis $\{
\ket{i} \}$ to be the simultaneous eigenbasis of the elements
$u(g)$, 
such that $u(g) \ket{i} = \chi_i(g) \ket{i}$, where each $\chi_i$ is a linear
representation of $G$.
Expressed in the basis $\{ \ket{i} \}$, \eqnref{mps_symmetry_condition} then becomes
\begin{equation}
V(g)^{\dagger} A[i] V(g) = \chi_i(g) A[i].
\end{equation}
Making use of \eqnref{h_chi_eqn}, we find that
\begin{equation}
V(g) \left\{V(h_{\chi_i})^{\dagger} A[i]\right\} = 
\left\{V(h_{\chi_i})^{\dagger} A[i]\right\} V(g).
\end{equation}
We can now conclude by Schur's Lemma that
\begin{equation}
A[i] = \widetilde{V}(h_{\chi_i}) \otimes A_{\mathrm{junk}}[i]
\end{equation}
for some operators $A_{\mathrm{junk}}[i]$. Therefore Theorem \ref{mainthm} holds with
$B_i = \widetilde{V}(h_{\chi_i})$. \qed

\emph{Non-trivial gates.}---In Theorem \ref{mainthm}, we have proven that the identity gate, which involves
measuring in the
simultaneous eigenbasis of the operators $u(g)$, is symmetry-protected. We will now see that
non-trivial gates (i.e.\ those involving measurement in a different basis) are not symmetry-protected.

For example, let us consider a
measurement that on the exact AKLT state would correspond to a rotation by an
angle $2 \theta$ about the $z$ axis (up to Pauli byproducts).
One of the possible measurement outcomes is $\ket{\theta}
\equiv \cos \theta \ket{x} + \sin \theta \ket{y}$.
Then from the decomposition (\ref{mps_tensor_decomposition_aklt}) of the
MPS tensor $A$ for a generic state in the Haldane phase, we find that
\begin{multline}
A[\theta] = (\cos \theta) \sigma_x \otimes A_{\mathrm{junk}}[x] + (\sin
\theta) \sigma_y \otimes A_{\mathrm{junk}}[y].
\end{multline}
If $A_{\mathrm{junk}}[x] = A_{\mathrm{junk}}[y]$ (as for the exact AKLT
state) then this implies
\begin{equation}
A[\theta] = [(\cos \theta) \sigma_x + (\sin \theta) \sigma_y] \otimes
A_{\mathrm{junk}}[x],
\end{equation}
and the evolution on the protected subsystem is the same as it would be for the
exact AKLT state.
However, there is no symmetry constraint that guarantees 
$A_{\mathrm{junk}}[x] = A_{\mathrm{junk}}[y]$ (because any choice whatsoever
for $A_{\mathrm{junk}}$ in \eqnref{mps_tensor_decomposition_aklt}
gives rise to an
MPS satisfying the symmetry constraints).
Therefore, the evolution induced by measurements in this basis is not fixed by
the symmetry; similar arguments apply to all non-trivial gates.

The preceding discussion of non-trivial gates applies to systems with
only the $Z_2 \times Z_2$ rotation symmetry, and larger symmetry groups will
lead to stronger constraints on the MPS tensor.
In particular, one might expect that for the AKLT state, imposing the full
$\mathrm{SO}(3)$ rotation symmetry would lead to all gates being protected, because
\emph{all} gates are achieved by measuring in the
basis $\{ \ket{x^{\prime}}, \ket{y^{\prime}}, \ket{z^{\prime}} \}$ for some
rotated orthogonal set of axes $x^{\prime},y^{\prime}, z^{\prime}$. This would
indeed be true if only the spin-1/2 projective representation $V_{1/2}(g)$ of
$\mathrm{SO}(3)$ appeared in the irrep
decomposition of $V(g)$, so that $V(g) = V_{1/2}(g) \otimes \mathbb{I}$. However, all the half-integer spin
representations of $\mathrm{SO}(3)$ have the same cohomology class, so this will not hold
in general.
Indeed, the numerical results of \cite{bartlett_renormalization} show reduced
performance of non-trivial gates. 
This should be contrasted with the protocol of \cite{miyake_edgestates}, where
a logical qubit is encoded into an explicitly spin-1/2 edge mode and particles are
adiabatically decoupled from the chain before being measured. In that case it
was found that all gates operate perfectly throughout the Haldane phase so long
as the full
rotational symmetry is maintained.

\emph{Initialization and readout.}---Apart from
performing unitary gates in correlation space, the other essential ingredient
for MBQC is the ability to initialize and read out the state of the
correlation system. It is easily verified (in the same way as for non-trivial
gates) that the usual procedures for doing this in the cluster or AKLT states are not
symmetry-protected. However,
a symmetry-protected readout \emph{can} be achieved throughout the Haldane phase by
terminating a finite chain of spin-1's with a spin-1/2, as in
\cite{bartlett_renormalization}.


\emph{Higher-dimensional systems.}---The notion of symmetry-protected
topological order has recently been extended to higher-dimensional systems
\cite{spto_2d,spto_higher}, and we speculate that our results could be generalized
in this context. However, if we consider a `quasi-1D' system whose
extent in all but one dimension is finite (but could be set arbitrarily large),
then the results of this Letter can be
applied directly.

\begin{figure}
\centering
\includegraphics[width=\linewidth]{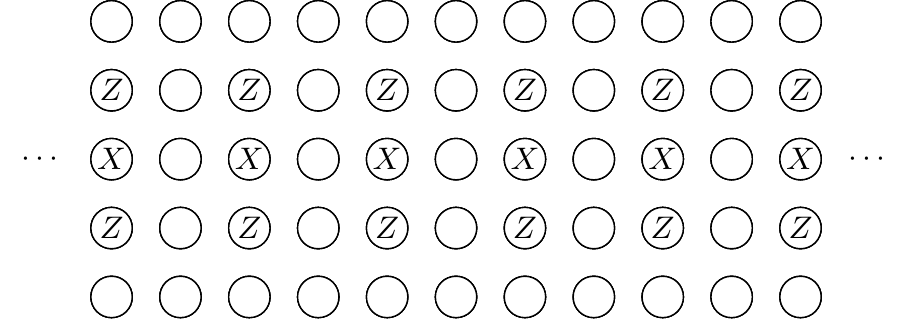}
\caption{\label{cluster_state_2d}
One generator of the $(Z_2 \times Z_2)^{\times N}$ symmetry in the 2-D cluster
state. The other
generators can be obtained from this one by a displacement by 1 horizontally
and/or an even number vertically. The circles represent qubits in the 2-D square
lattice.}
\end{figure}

For example, a \twod{} cluster model of extent $2N$ in the
vertical direction (with periodic boundary conditions in that direction) is in a
non-trivial symmetry-protected phase with respect to the $(Z_2 \times Z_2)^{\times N}$ symmetry depicted in
Figure \ref{cluster_state_2d}. This symmetry is represented in correlation space by a
tensor product of $N$ copies of the Pauli representation; this is a maximally
non-commutative projective representation of the symmetry group. By Lemma
\ref{uniqueness_lemma}, the protected subsystem has dimension $2^N$.
Therefore, throughout the symmetry-protected phase
there is a capacity for $N$ qubits to be propagated in the horizontal direction by
measuring each `site' (here a pair of adjacent columns) in a simultaneous eigenbasis of the
symmetry. For the particular representation of $(Z_2 \times Z_2)^{\times N}$ 
depicted in Figure \ref{cluster_state_2d}, it
is straightforward to show that there exists such an eigenbasis which is also a
product basis over the qubits making up the site, so that this propagation can
be achieved by single-qubit measurements.

\emph{Conclusion.---}In summary, we have identified a class of
symmetry-protected topological orders, each of which ensures the perfect operation of the
identity gate in MBQC throughout an entire symmetry-protected phase. Such
connections between MBQC and quantum order can be expected to lead to a greater
understanding of the potential for single-particle measurements on ground states
of quantum spin systems to be a robust form of quantum computation.

By contrast,
we have shown that the perfect operation of non-trivial gates is a
property only of specific systems within such a phase, contrary to some previous
hopes \cite{doherty-bartlett-prl-2008}. However, we have not
given a complete characterization of the operation of non-trivial gates away
from these points, and it is possible that their performance could be made
arbitrarily good by a suitable choice of adaptive measurement protocol, as in
\cite{bartlett_renormalization}.

\begin{acknowledgments}
We acknowledge discussions with A.\ Miyake, and support from the ARC via the
Centre of Excellence in Engineered Quantum Systems (EQuS), project number
CE110001013. I.S.\ acknowledges support from the Australia-Israel Scientific
Exchange Foundation (AISEF).
\end{acknowledgments}


\begin{thebibliography}{32}%
\makeatletter
\providecommand \@ifxundefined [1]{%
 \@ifx{#1\undefined}
}%
\providecommand \@ifnum [1]{%
 \ifnum #1\expandafter \@firstoftwo
 \else \expandafter \@secondoftwo
 \fi
}%
\providecommand \@ifx [1]{%
 \ifx #1\expandafter \@firstoftwo
 \else \expandafter \@secondoftwo
 \fi
}%
\providecommand \natexlab [1]{#1}%
\providecommand \enquote  [1]{``#1''}%
\providecommand \bibnamefont  [1]{#1}%
\providecommand \bibfnamefont [1]{#1}%
\providecommand \citenamefont [1]{#1}%
\providecommand \href@noop [0]{\@secondoftwo}%
\providecommand \href [0]{\begingroup \@sanitize@url \@href}%
\providecommand \@href[1]{\@@startlink{#1}\@@href}%
\providecommand \@@href[1]{\endgroup#1\@@endlink}%
\providecommand \@sanitize@url [0]{\catcode `\\12\catcode `\$12\catcode
  `\&12\catcode `\#12\catcode `\^12\catcode `\_12\catcode `\%12\relax}%
\providecommand \@@startlink[1]{}%
\providecommand \@@endlink[0]{}%
\providecommand \url  [0]{\begingroup\@sanitize@url \@url }%
\providecommand \@url [1]{\endgroup\@href {#1}{\urlprefix }}%
\providecommand \urlprefix  [0]{URL }%
\providecommand \Eprint [0]{\href }%
\@ifxundefined \urlstyle {%
  \providecommand \doi  [0]{\begingroup \@sanitize@url \@doi}%
  \providecommand \@doi [1]{\endgroup \@@startlink {\doibase
  #1}doi:\discretionary {}{}{}#1\@@endlink }%
}{%
  \providecommand \doi  [0]{doi:\discretionary{}{}{}\begingroup
  \urlstyle{rm}\Url }%
}%
\providecommand \doibase [0]{http://dx.doi.org/}%
\providecommand \Doi [0]{\begingroup \@sanitize@url \@Doi }%
\providecommand \@Doi  [1]{\endgroup\@@startlink{\doibase#1}\@@Doi}%
\providecommand \@@Doi [1]{#1\@@endlink}%
\providecommand \selectlanguage [0]{\@gobble}%
\providecommand \bibinfo  [0]{\@secondoftwo}%
\providecommand \bibfield  [0]{\@secondoftwo}%
\providecommand \translation [1]{[#1]}%
\providecommand \BibitemOpen [0]{}%
\providecommand \bibitemStop [0]{}%
\providecommand \bibitemNoStop [0]{.\EOS\space}%
\providecommand \EOS [0]{\spacefactor3000\relax}%
\providecommand \BibitemShut  [1]{\csname bibitem#1\endcsname}%
\bibitem [{\citenamefont {Raussendorf}\ and\ \citenamefont
  {Briegel}(2001)}]{raussendorf-prl-2001}%
  \BibitemOpen
  \bibfield  {author} {\bibinfo {author} {\bibfnamefont {Robert}\ \bibnamefont
  {Raussendorf}}\ and\ \bibinfo {author} {\bibfnamefont {Hans~J.}\ \bibnamefont
  {Briegel}},\ }\bibfield  {title} {\enquote {\bibinfo {title} {A one-way
  quantum computer},}\ }\Doi {10.1103/PhysRevLett.86.5188} {\bibfield
  {journal} {\bibinfo  {journal} {Phys. Rev. Lett.},\ }\textbf {\bibinfo
  {volume} {86}},\ \bibinfo {pages} {5188} (\bibinfo {year}
  {2001})}\BibitemShut {NoStop}%
\bibitem [{\citenamefont {Raussendorf}\ \emph {et~al.}(2003)\citenamefont
  {Raussendorf}, \citenamefont {Browne},\ and\ \citenamefont
  {Briegel}}]{raussendorf_et_al_2003}%
  \BibitemOpen
  \bibfield  {author} {\bibinfo {author} {\bibfnamefont {Robert}\ \bibnamefont
  {Raussendorf}}, \bibinfo {author} {\bibfnamefont {Daniel~E.}\ \bibnamefont
  {Browne}}, \ and\ \bibinfo {author} {\bibfnamefont {Hans~J.}\ \bibnamefont
  {Briegel}},\ }\bibfield  {title} {\enquote {\bibinfo {title}
  {Measurement-based quantum computation on cluster states},}\ }\Doi
  {10.1103/PhysRevA.68.022312} {\bibfield  {journal} {\bibinfo  {journal}
  {Phys. Rev. A},\ }\textbf {\bibinfo {volume} {68}},\ \bibinfo {pages}
  {022312} (\bibinfo {year} {2003})},\ \Eprint
  {http://arxiv.org/abs/arXiv:quant-ph/0301052} {arXiv:quant-ph/0301052}
  \BibitemShut {NoStop}%
\bibitem [{\citenamefont {Briegel}\ \emph {et~al.}(2009)\citenamefont
  {Briegel}, \citenamefont {Browne}, \citenamefont {Dur}, \citenamefont
  {Raussendorf},\ and\ \citenamefont {Van~den Nest}}]{briegel_mbqc_natphys}%
  \BibitemOpen
  \bibfield  {author} {\bibinfo {author} {\bibfnamefont {H.~J.}\ \bibnamefont
  {Briegel}}, \bibinfo {author} {\bibfnamefont {D.~E.}\ \bibnamefont {Browne}},
  \bibinfo {author} {\bibfnamefont {W.}~\bibnamefont {Dur}}, \bibinfo {author}
  {\bibfnamefont {R.}~\bibnamefont {Raussendorf}}, \ and\ \bibinfo {author}
  {\bibfnamefont {M.}~\bibnamefont {Van~den Nest}},\ }\bibfield  {title}
  {\enquote {\bibinfo {title} {Measurement-based quantum computation},}\ }\Doi
  {10.1038/nphys1157} {\bibfield  {journal} {\bibinfo  {journal} {Nat. Phys.},\
  }\textbf {\bibinfo {volume} {5}},\ \bibinfo {pages} {19} (\bibinfo {year}
  {2009})},\ \Eprint {http://arxiv.org/abs/0910.1116} {arXiv:0910.1116}
  \BibitemShut {NoStop}%
\bibitem [{\citenamefont {Doherty}\ and\ \citenamefont
  {Bartlett}(2009)}]{doherty-bartlett-prl-2008}%
  \BibitemOpen
  \bibfield  {author} {\bibinfo {author} {\bibfnamefont {Andrew~C.}\
  \bibnamefont {Doherty}}\ and\ \bibinfo {author} {\bibfnamefont {Stephen~D.}\
  \bibnamefont {Bartlett}},\ }\bibfield  {title} {\enquote {\bibinfo {title}
  {Identifying phases of quantum many-body systems that are universal for
  quantum computation},}\ }\Doi {10.1103/PhysRevLett.103.020506} {\bibfield
  {journal} {\bibinfo  {journal} {Phys. Rev. Lett.},\ }\textbf {\bibinfo
  {volume} {103}},\ \bibinfo {pages} {020506} (\bibinfo {year} {2009})},\
  \Eprint {http://arxiv.org/abs/arXiv:0802.4314} {arXiv:0802.4314} \BibitemShut
  {NoStop}%
\bibitem [{\citenamefont {Chen}\ \emph {et~al.}(2009)\citenamefont {Chen},
  \citenamefont {Zeng}, \citenamefont {Gu}, \citenamefont {Yoshida},\ and\
  \citenamefont {Chuang}}]{chen_tricluster}%
  \BibitemOpen
  \bibfield  {author} {\bibinfo {author} {\bibfnamefont {Xie}\ \bibnamefont
  {Chen}}, \bibinfo {author} {\bibfnamefont {Bei}\ \bibnamefont {Zeng}},
  \bibinfo {author} {\bibfnamefont {Zheng-Cheng}\ \bibnamefont {Gu}}, \bibinfo
  {author} {\bibfnamefont {Beni}\ \bibnamefont {Yoshida}}, \ and\ \bibinfo
  {author} {\bibfnamefont {Isaac~L.}\ \bibnamefont {Chuang}},\ }\bibfield
  {title} {\enquote {\bibinfo {title} {Gapped two-body hamiltonian whose unique
  ground state is universal for one-way quantum computation},}\ }\Doi
  {10.1103/PhysRevLett.102.220501} {\bibfield  {journal} {\bibinfo  {journal}
  {Phys. Rev. Lett.},\ }\textbf {\bibinfo {volume} {102}},\ \bibinfo {pages}
  {220501} (\bibinfo {year} {2009})},\ \Eprint {http://arxiv.org/abs/0812.4067}
  {arXiv:0812.4067} \BibitemShut {NoStop}%
\bibitem [{\citenamefont {Miyake}(2011)}]{miyake_aklt}%
  \BibitemOpen
  \bibfield  {author} {\bibinfo {author} {\bibfnamefont {Akimasa}\ \bibnamefont
  {Miyake}},\ }\bibfield  {title} {\enquote {\bibinfo {title} {Quantum
  computational capability of a {2D} valence bond solid phase},}\ }\Doi
  {10.1016/j.aop.2011.03.006} {\bibfield  {journal} {\bibinfo  {journal} {Ann.
  Phys.},\ }\textbf {\bibinfo {volume} {326}},\ \bibinfo {pages} {1656}
  (\bibinfo {year} {2011})},\ \Eprint {http://arxiv.org/abs/1009.3491}
  {arXiv:1009.3491} \BibitemShut {NoStop}%
\bibitem [{\citenamefont {Wei}\ \emph {et~al.}(2011)\citenamefont {Wei},
  \citenamefont {Affleck},\ and\ \citenamefont
  {Raussendorf}}]{raussendorf_aklt1}%
  \BibitemOpen
  \bibfield  {author} {\bibinfo {author} {\bibfnamefont {Tzu-Chieh}\
  \bibnamefont {Wei}}, \bibinfo {author} {\bibfnamefont {Ian}\ \bibnamefont
  {Affleck}}, \ and\ \bibinfo {author} {\bibfnamefont {Robert}\ \bibnamefont
  {Raussendorf}},\ }\bibfield  {title} {\enquote {\bibinfo {title}
  {{Affleck-Kennedy-Lieb-Tasaki} state on a honeycomb lattice is a universal
  quantum computational resource},}\ }\Doi {10.1103/PhysRevLett.106.070501}
  {\bibfield  {journal} {\bibinfo  {journal} {Phys. Rev. Lett.},\ }\textbf
  {\bibinfo {volume} {106}},\ \bibinfo {pages} {070501} (\bibinfo {year}
  {2011})},\ \Eprint {http://arxiv.org/abs/1102.5064} {arXiv:1102.5064}
  \BibitemShut {NoStop}%
\bibitem [{\citenamefont {Gu}\ and\ \citenamefont {Wen}(2009)}]{gu_wen_2009}%
  \BibitemOpen
  \bibfield  {author} {\bibinfo {author} {\bibfnamefont {Zheng-Cheng}\
  \bibnamefont {Gu}}\ and\ \bibinfo {author} {\bibfnamefont {Xiao-Gang}\
  \bibnamefont {Wen}},\ }\bibfield  {title} {\enquote {\bibinfo {title}
  {Tensor-entanglement-filtering renormalization approach and
  symmetry-protected topological order},}\ }\Doi {10.1103/PhysRevB.80.155131}
  {\bibfield  {journal} {\bibinfo  {journal} {Phys. Rev. B},\ }\textbf
  {\bibinfo {volume} {80}},\ \bibinfo {pages} {155131} (\bibinfo {year}
  {2009})},\ \Eprint {http://arxiv.org/abs/0903.1069} {arXiv:0903.1069}
  \BibitemShut {NoStop}%
\bibitem [{\citenamefont {Chen}\ \emph
  {et~al.}(2011){\natexlab{a}}\citenamefont {Chen}, \citenamefont {Gu},\ and\
  \citenamefont {Wen}}]{chen_gu_wen}%
  \BibitemOpen
  \bibfield  {author} {\bibinfo {author} {\bibfnamefont {Xie}\ \bibnamefont
  {Chen}}, \bibinfo {author} {\bibfnamefont {Zheng-Cheng}\ \bibnamefont {Gu}},
  \ and\ \bibinfo {author} {\bibfnamefont {Xiao-Gang}\ \bibnamefont {Wen}},\
  }\bibfield  {title} {\enquote {\bibinfo {title} {Classification of gapped
  symmetric phases in one-dimensional spin systems},}\ }\Doi
  {10.1103/PhysRevB.83.035107} {\bibfield  {journal} {\bibinfo  {journal}
  {Phys. Rev. B},\ }\textbf {\bibinfo {volume} {83}},\ \bibinfo {pages}
  {035107} (\bibinfo {year} {2011}{\natexlab{a}})},\ \Eprint
  {http://arxiv.org/abs/arXiv:1008.3745} {arXiv:1008.3745} \BibitemShut
  {NoStop}%
\bibitem [{\citenamefont {Schuch}\ \emph {et~al.}(2011)\citenamefont {Schuch},
  \citenamefont {P{\'e}rez-Garc{\'i}a},\ and\ \citenamefont {Cirac}}]{schuch}%
  \BibitemOpen
  \bibfield  {author} {\bibinfo {author} {\bibfnamefont {Norbert}\ \bibnamefont
  {Schuch}}, \bibinfo {author} {\bibfnamefont {David}\ \bibnamefont
  {P{\'e}rez-Garc{\'i}a}}, \ and\ \bibinfo {author} {\bibfnamefont {Ignacio}\
  \bibnamefont {Cirac}},\ }\bibfield  {title} {\enquote {\bibinfo {title}
  {Classifying quantum phases using matrix product states and projected
  entangled pair states},}\ }\Doi {10.1103/PhysRevB.84.165139} {\bibfield
  {journal} {\bibinfo  {journal} {Phys. Rev. B},\ }\textbf {\bibinfo {volume}
  {84}},\ \bibinfo {pages} {165139} (\bibinfo {year} {2011})},\ \Eprint
  {http://arxiv.org/abs/1010.3732} {arXiv:1010.3732} \BibitemShut {NoStop}%
\bibitem [{\citenamefont {Gross}\ and\ \citenamefont
  {Eisert}(2010)}]{gross-eisert-2010-webs}%
  \BibitemOpen
  \bibfield  {author} {\bibinfo {author} {\bibfnamefont {D.}~\bibnamefont
  {Gross}}\ and\ \bibinfo {author} {\bibfnamefont {J.}~\bibnamefont {Eisert}},\
  }\bibfield  {title} {\enquote {\bibinfo {title} {Quantum computational
  webs},}\ }\Doi {10.1103/PhysRevA.82.040303} {\bibfield  {journal} {\bibinfo
  {journal} {Phys. Rev. A},\ }\textbf {\bibinfo {volume} {82}},\ \bibinfo
  {pages} {040303} (\bibinfo {year} {2010})},\ \Eprint
  {http://arxiv.org/abs/arXiv:0810.2542} {arXiv:0810.2542} \BibitemShut
  {NoStop}%
\bibitem [{\citenamefont {Affleck}\ \emph {et~al.}(1987)\citenamefont
  {Affleck}, \citenamefont {Kennedy}, \citenamefont {Lieb},\ and\ \citenamefont
  {Tasaki}}]{aklt2}%
  \BibitemOpen
  \bibfield  {author} {\bibinfo {author} {\bibfnamefont {Ian}\ \bibnamefont
  {Affleck}}, \bibinfo {author} {\bibfnamefont {Tom}\ \bibnamefont {Kennedy}},
  \bibinfo {author} {\bibfnamefont {Elliott~H.}\ \bibnamefont {Lieb}}, \ and\
  \bibinfo {author} {\bibfnamefont {Hal}\ \bibnamefont {Tasaki}},\ }\bibfield
  {title} {\enquote {\bibinfo {title} {Rigorous results on valence-bond ground
  states in antiferromagnets},}\ }\Doi {10.1103/PhysRevLett.59.799} {\bibfield
  {journal} {\bibinfo  {journal} {Phys. Rev. Lett.},\ }\textbf {\bibinfo
  {volume} {59}},\ \bibinfo {pages} {799} (\bibinfo {year} {1987})}\BibitemShut
  {NoStop}%
\bibitem [{\citenamefont {Affleck}\ \emph {et~al.}(1988)\citenamefont
  {Affleck}, \citenamefont {Kennedy}, \citenamefont {Lieb},\ and\ \citenamefont
  {Tasaki}}]{aklt1}%
  \BibitemOpen
  \bibfield  {author} {\bibinfo {author} {\bibfnamefont {I.}~\bibnamefont
  {Affleck}}, \bibinfo {author} {\bibfnamefont {T.}~\bibnamefont {Kennedy}},
  \bibinfo {author} {\bibfnamefont {E.H.}\ \bibnamefont {Lieb}}, \ and\
  \bibinfo {author} {\bibfnamefont {H.}~\bibnamefont {Tasaki}},\ }\bibfield
  {title} {\enquote {\bibinfo {title} {Valence bond ground states in isotropic
  quantum antiferromagnets},}\ }\Doi {10.1007/BF01218021} {\bibfield  {journal}
  {\bibinfo  {journal} {Commun. Math. Phys.},\ }\textbf {\bibinfo {volume}
  {115}},\ \bibinfo {pages} {477} (\bibinfo {year} {1988})}\BibitemShut
  {NoStop}%
\bibitem [{\citenamefont {Brennen}\ and\ \citenamefont
  {Miyake}(2008)}]{brennen-aklt-mbqc-2008}%
  \BibitemOpen
  \bibfield  {author} {\bibinfo {author} {\bibfnamefont {Gavin~K.}\
  \bibnamefont {Brennen}}\ and\ \bibinfo {author} {\bibfnamefont {Akimasa}\
  \bibnamefont {Miyake}},\ }\bibfield  {title} {\enquote {\bibinfo {title}
  {Measurement-based quantum computer in the gapped ground state of a two-body
  hamiltonian},}\ }\Doi {10.1103/PhysRevLett.101.010502} {\bibfield  {journal}
  {\bibinfo  {journal} {Phys. Rev. Lett.},\ }\textbf {\bibinfo {volume}
  {101}},\ \bibinfo {pages} {010502} (\bibinfo {year} {2008})},\ \Eprint
  {http://arxiv.org/abs/arXiv:0803.1478} {arXiv:0803.1478} \BibitemShut
  {NoStop}%
\bibitem [{\citenamefont {Pollmann}\ \emph {et~al.}(2012)\citenamefont
  {Pollmann}, \citenamefont {Berg}, \citenamefont {Turner},\ and\ \citenamefont
  {Oshikawa}}]{pollmann-arxiv-2009}%
  \BibitemOpen
  \bibfield  {author} {\bibinfo {author} {\bibfnamefont {Frank}\ \bibnamefont
  {Pollmann}}, \bibinfo {author} {\bibfnamefont {Erez}\ \bibnamefont {Berg}},
  \bibinfo {author} {\bibfnamefont {Ari~M.}\ \bibnamefont {Turner}}, \ and\
  \bibinfo {author} {\bibfnamefont {Masaki}\ \bibnamefont {Oshikawa}},\
  }\bibfield  {title} {\enquote {\bibinfo {title} {Symmetry protection of
  topological phases in one-dimensional quantum spin systems},}\ }\Doi
  {10.1103/PhysRevB.85.075125} {\bibfield  {journal} {\bibinfo  {journal}
  {Phys. Rev. B},\ }\textbf {\bibinfo {volume} {85}},\ \bibinfo {pages}
  {075125} (\bibinfo {year} {2012})},\ \Eprint {http://arxiv.org/abs/0909.4059}
  {arXiv:0909.4059} \BibitemShut {NoStop}%
\bibitem [{\citenamefont {Pollmann}\ \emph {et~al.}(2010)\citenamefont
  {Pollmann}, \citenamefont {Turner}, \citenamefont {Berg},\ and\ \citenamefont
  {Oshikawa}}]{pollmann-prb-2010}%
  \BibitemOpen
  \bibfield  {author} {\bibinfo {author} {\bibfnamefont {Frank}\ \bibnamefont
  {Pollmann}}, \bibinfo {author} {\bibfnamefont {Ari~M.}\ \bibnamefont
  {Turner}}, \bibinfo {author} {\bibfnamefont {Erez}\ \bibnamefont {Berg}}, \
  and\ \bibinfo {author} {\bibfnamefont {Masaki}\ \bibnamefont {Oshikawa}},\
  }\bibfield  {title} {\enquote {\bibinfo {title} {Entanglement spectrum of a
  topological phase in one dimension},}\ }\Doi {10.1103/PhysRevB.81.064439}
  {\bibfield  {journal} {\bibinfo  {journal} {Phys. Rev. B},\ }\textbf
  {\bibinfo {volume} {81}},\ \bibinfo {pages} {064439} (\bibinfo {year}
  {2010})},\ \Eprint {http://arxiv.org/abs/arXiv:0910.1811} {arXiv:0910.1811}
  \BibitemShut {NoStop}%
\bibitem [{\citenamefont {Barjaktarevic}\ \emph {et~al.}(2005)\citenamefont
  {Barjaktarevic}, \citenamefont {McKenzie}, \citenamefont {Links},\ and\
  \citenamefont {Milburn}}]{uq_teleportation}%
  \BibitemOpen
  \bibfield  {author} {\bibinfo {author} {\bibfnamefont {J.~P.}\ \bibnamefont
  {Barjaktarevic}}, \bibinfo {author} {\bibfnamefont {R.~H.}\ \bibnamefont
  {McKenzie}}, \bibinfo {author} {\bibfnamefont {J.}~\bibnamefont {Links}}, \
  and\ \bibinfo {author} {\bibfnamefont {G.~J.}\ \bibnamefont {Milburn}},\
  }\bibfield  {title} {\enquote {\bibinfo {title} {Measurement-based
  teleportation along quantum spin chains},}\ }\Doi
  {10.1103/PhysRevLett.95.230501} {\bibfield  {journal} {\bibinfo  {journal}
  {Phys. Rev. Lett.},\ }\textbf {\bibinfo {volume} {95}},\ \bibinfo {pages}
  {230501} (\bibinfo {year} {2005})},\ \Eprint
  {http://arxiv.org/abs/quant-ph/0501180} {arXiv:quant-ph/0501180} \BibitemShut
  {NoStop}%
\bibitem [{\citenamefont {Bartlett}\ \emph {et~al.}(2010)\citenamefont
  {Bartlett}, \citenamefont {Brennen}, \citenamefont {Miyake},\ and\
  \citenamefont {Renes}}]{bartlett_renormalization}%
  \BibitemOpen
  \bibfield  {author} {\bibinfo {author} {\bibfnamefont {Stephen~D.}\
  \bibnamefont {Bartlett}}, \bibinfo {author} {\bibfnamefont {Gavin~K.}\
  \bibnamefont {Brennen}}, \bibinfo {author} {\bibfnamefont {Akimasa}\
  \bibnamefont {Miyake}}, \ and\ \bibinfo {author} {\bibfnamefont {Joseph~M.}\
  \bibnamefont {Renes}},\ }\bibfield  {title} {\enquote {\bibinfo {title}
  {Quantum computational renormalization in the {Haldane} phase},}\ }\Doi
  {10.1103/PhysRevLett.105.110502} {\bibfield  {journal} {\bibinfo  {journal}
  {Phys. Rev. Lett.},\ }\textbf {\bibinfo {volume} {105}},\ \bibinfo {pages}
  {110502} (\bibinfo {year} {2010})},\ \Eprint
  {http://arxiv.org/abs/arXiv:1004.4906} {arXiv:1004.4906} \BibitemShut
  {NoStop}%
\bibitem [{\citenamefont {Venuti}\ and\ \citenamefont
  {Roncaglia}(2005)}]{venuti-2005-le-string}%
  \BibitemOpen
  \bibfield  {author} {\bibinfo {author} {\bibfnamefont {L.C.}\ \bibnamefont
  {Venuti}}\ and\ \bibinfo {author} {\bibfnamefont {M.}~\bibnamefont
  {Roncaglia}},\ }\bibfield  {title} {\enquote {\bibinfo {title} {Analytic
  relations between localizable entanglement and string correlations in spin
  systems},}\ }\Doi {10.1103/PhysRevLett.94.207207} {\bibfield  {journal}
  {\bibinfo  {journal} {Phys. Rev. Lett.},\ }\textbf {\bibinfo {volume} {94}},\
  \bibinfo {pages} {207207} (\bibinfo {year} {2005})},\ \Eprint
  {http://arxiv.org/abs/arXiv:cond-mat/0503021} {arXiv:cond-mat/0503021}
  \BibitemShut {NoStop}%
\bibitem [{\citenamefont {Zhou}\ \emph {et~al.}(2003)\citenamefont {Zhou},
  \citenamefont {Zeng}, \citenamefont {Xu},\ and\ \citenamefont
  {Sun}}]{qudit_cluster_state}%
  \BibitemOpen
  \bibfield  {author} {\bibinfo {author} {\bibfnamefont {D.~L.}\ \bibnamefont
  {Zhou}}, \bibinfo {author} {\bibfnamefont {B.}~\bibnamefont {Zeng}}, \bibinfo
  {author} {\bibfnamefont {Z.}~\bibnamefont {Xu}}, \ and\ \bibinfo {author}
  {\bibfnamefont {C.~P.}\ \bibnamefont {Sun}},\ }\bibfield  {title} {\enquote
  {\bibinfo {title} {Quantum computation based on \textit{d}-level cluster
  state},}\ }\Doi {10.1103/PhysRevA.68.062303} {\bibfield  {journal} {\bibinfo
  {journal} {Phys. Rev. A},\ }\textbf {\bibinfo {volume} {68}},\ \bibinfo
  {pages} {062303} (\bibinfo {year} {2003})},\ \Eprint
  {http://arxiv.org/abs/quant-ph/0304054} {arXiv:quant-ph/0304054} \BibitemShut
  {NoStop}%
\bibitem [{\citenamefont {Gross}\ \emph {et~al.}(2007)\citenamefont {Gross},
  \citenamefont {Eisert}, \citenamefont {Schuch},\ and\ \citenamefont
  {Perez-Garcia}}]{correlation_space_pra}%
  \BibitemOpen
  \bibfield  {author} {\bibinfo {author} {\bibfnamefont {D.}~\bibnamefont
  {Gross}}, \bibinfo {author} {\bibfnamefont {J.}~\bibnamefont {Eisert}},
  \bibinfo {author} {\bibfnamefont {N.}~\bibnamefont {Schuch}}, \ and\ \bibinfo
  {author} {\bibfnamefont {D.}~\bibnamefont {Perez-Garcia}},\ }\bibfield
  {title} {\enquote {\bibinfo {title} {Measurement-based quantum computation
  beyond the one-way model},}\ }\Doi {10.1103/PhysRevA.76.052315} {\bibfield
  {journal} {\bibinfo  {journal} {Phys. Rev. A},\ }\textbf {\bibinfo {volume}
  {76}},\ \bibinfo {pages} {052315} (\bibinfo {year} {2007})},\ \Eprint
  {http://arxiv.org/abs/arXiv:0706.3401} {arXiv:0706.3401} \BibitemShut
  {NoStop}%
\bibitem [{\citenamefont {Verstraete}\ and\ \citenamefont
  {Cirac}(2006)}]{mps_faithfully}%
  \BibitemOpen
  \bibfield  {author} {\bibinfo {author} {\bibfnamefont {F.}~\bibnamefont
  {Verstraete}}\ and\ \bibinfo {author} {\bibfnamefont {J.~I.}\ \bibnamefont
  {Cirac}},\ }\bibfield  {title} {\enquote {\bibinfo {title} {Matrix product
  states represent ground states faithfully},}\ }\Doi
  {10.1103/PhysRevB.73.094423} {\bibfield  {journal} {\bibinfo  {journal}
  {Phys. Rev. B},\ }\textbf {\bibinfo {volume} {73}},\ \bibinfo {pages}
  {094423} (\bibinfo {year} {2006})},\ \Eprint
  {http://arxiv.org/abs/arXiv:cond-mat/0505140} {arXiv:cond-mat/0505140}
  \BibitemShut {NoStop}%
\bibitem [{\citenamefont {Hastings}(2007)}]{hastings_area_law}%
  \BibitemOpen
  \bibfield  {author} {\bibinfo {author} {\bibfnamefont {M.~B.}\ \bibnamefont
  {Hastings}},\ }\bibfield  {title} {\enquote {\bibinfo {title} {An area law
  for one-dimensional quantum systems},}\ }\Doi {10.1088/1742/2007/08/P08024}
  {\bibfield  {journal} {\bibinfo  {journal} {Journal of Statistical Mechanics:
  Theory and Experiment},\ }\textbf {\bibinfo {volume} {2007}},\ \bibinfo
  {pages} {P08024} (\bibinfo {year} {2007})},\ \Eprint
  {http://arxiv.org/abs/arXiv:0705.2024} {arXiv:0705.2024} \BibitemShut
  {NoStop}%
\bibitem [{\citenamefont {P\'erez-Garc\'\i{}a}\ \emph
  {et~al.}(2008)\citenamefont {P\'erez-Garc\'\i{}a}, \citenamefont {Wolf},
  \citenamefont {Sanz}, \citenamefont {Verstraete},\ and\ \citenamefont
  {Cirac}}]{string_order_symmetries}%
  \BibitemOpen
  \bibfield  {author} {\bibinfo {author} {\bibfnamefont {D.}~\bibnamefont
  {P\'erez-Garc\'\i{}a}}, \bibinfo {author} {\bibfnamefont {M.~M.}\
  \bibnamefont {Wolf}}, \bibinfo {author} {\bibfnamefont {M.}~\bibnamefont
  {Sanz}}, \bibinfo {author} {\bibfnamefont {F.}~\bibnamefont {Verstraete}}, \
  and\ \bibinfo {author} {\bibfnamefont {J.~I.}\ \bibnamefont {Cirac}},\
  }\bibfield  {title} {\enquote {\bibinfo {title} {String order and symmetries
  in quantum spin lattices},}\ }\Doi {10.1103/PhysRevLett.100.167202}
  {\bibfield  {journal} {\bibinfo  {journal} {Phys. Rev. Lett.},\ }\textbf
  {\bibinfo {volume} {100}},\ \bibinfo {pages} {167202} (\bibinfo {year}
  {2008})},\ \Eprint {http://arxiv.org/abs/arXiv:0802.0447} {arXiv:0802.0447}
  \BibitemShut {NoStop}%
\bibitem [{\citenamefont {Singh}\ \emph {et~al.}(2010)\citenamefont {Singh},
  \citenamefont {Pfeifer},\ and\ \citenamefont
  {Vidal}}]{tensor_network_global_symmetry}%
  \BibitemOpen
  \bibfield  {author} {\bibinfo {author} {\bibfnamefont {Sukhwinder}\
  \bibnamefont {Singh}}, \bibinfo {author} {\bibfnamefont {Robert N.~C.}\
  \bibnamefont {Pfeifer}}, \ and\ \bibinfo {author} {\bibfnamefont {Guifr\'e}\
  \bibnamefont {Vidal}},\ }\bibfield  {title} {\enquote {\bibinfo {title}
  {Tensor network decompositions in the presence of a global symmetry},}\ }\Doi
  {10.1103/PhysRevA.82.050301} {\bibfield  {journal} {\bibinfo  {journal}
  {Phys. Rev. A},\ }\textbf {\bibinfo {volume} {82}},\ \bibinfo {pages}
  {050301} (\bibinfo {year} {2010})},\ \Eprint {http://arxiv.org/abs/0907.2994}
  {arXiv:0907.2994} \BibitemShut {NoStop}%
\bibitem [{\citenamefont {Son}\ \emph {et~al.}(2011)\citenamefont {Son},
  \citenamefont {Amico}, \citenamefont {Fazio}, \citenamefont {Hamma},
  \citenamefont {Pascazio},\ and\ \citenamefont {Vedral}}]{son-cluster-2011}%
  \BibitemOpen
  \bibfield  {author} {\bibinfo {author} {\bibfnamefont {W.}~\bibnamefont
  {Son}}, \bibinfo {author} {\bibfnamefont {L.}~\bibnamefont {Amico}}, \bibinfo
  {author} {\bibfnamefont {R.}~\bibnamefont {Fazio}}, \bibinfo {author}
  {\bibfnamefont {A.}~\bibnamefont {Hamma}}, \bibinfo {author} {\bibfnamefont
  {S.}~\bibnamefont {Pascazio}}, \ and\ \bibinfo {author} {\bibfnamefont
  {V.}~\bibnamefont {Vedral}},\ }\bibfield  {title} {\enquote {\bibinfo {title}
  {Quantum phase transition between cluster and antiferromagnetic states},}\
  }\Doi {10.1209/0295-5075/95/50001} {\bibfield  {journal} {\bibinfo  {journal}
  {Europhys. Lett.},\ }\textbf {\bibinfo {volume} {95}},\ \bibinfo {pages}
  {50001} (\bibinfo {year} {2011})},\ \Eprint {http://arxiv.org/abs/1103.0251}
  {arXiv:1103.0251} \BibitemShut {NoStop}%
\bibitem [{\citenamefont {Kleppner}(1965)}]{kleppner_multipliers}%
  \BibitemOpen
  \bibfield  {author} {\bibinfo {author} {\bibfnamefont {Adam}\ \bibnamefont
  {Kleppner}},\ }\bibfield  {title} {\enquote {\bibinfo {title} {Multipliers on
  abelian groups},}\ }\Doi {10.1007/BF01370393} {\bibfield  {journal} {\bibinfo
   {journal} {Mathematische Annalen},\ }\textbf {\bibinfo {volume} {158}},\
  \bibinfo {pages} {11} (\bibinfo {year} {1965})}\BibitemShut {NoStop}%
\bibitem [{\citenamefont {Frucht}(1932)}]{frucht1932}%
  \BibitemOpen
  \bibfield  {author} {\bibinfo {author} {\bibfnamefont {R.}~\bibnamefont
  {Frucht}},\ }\bibfield  {title} {\enquote {\bibinfo {title} {{\"U}ber die
  darstellung endlicher abelscher gruppen durch kollineationen},}\ }\Doi
  {10.1515/crll.1932.166.16} {\bibfield  {journal} {\bibinfo  {journal}
  {Journal f\"ur die reine und angewandte Mathematik},\ }\textbf {\bibinfo
  {volume} {1932}},\ \bibinfo {pages} {16} (\bibinfo {year}
  {1932})}\BibitemShut {NoStop}%
\bibitem [{\citenamefont {Berkovich}\ and\ \citenamefont
  {Zhmud$^\prime$}(1998)}]{berkovich_characters}%
  \BibitemOpen
  \bibfield  {author} {\bibinfo {author} {\bibfnamefont {Ya.~G.}\ \bibnamefont
  {Berkovich}}\ and\ \bibinfo {author} {\bibfnamefont {E.~M.}\ \bibnamefont
  {Zhmud$^\prime$}},\ }\href@noop {} {\emph {\bibinfo {title} {Characters of
  Finite Groups}}},\ Vol.~\bibinfo {volume} {1}\ (\bibinfo  {publisher}
  {American Mathematical Society},\ \bibinfo {address} {Providence, Rhode
  Island},\ \bibinfo {year} {1998})\BibitemShut {NoStop}%
\bibitem [{\citenamefont {Miyake}(2010)}]{miyake_edgestates}%
  \BibitemOpen
  \bibfield  {author} {\bibinfo {author} {\bibfnamefont {Akimasa}\ \bibnamefont
  {Miyake}},\ }\bibfield  {title} {\enquote {\bibinfo {title} {Quantum
  computation on the edge of a symmetry-protected topological order},}\ }\Doi
  {10.1103/PhysRevLett.105.040501} {\bibfield  {journal} {\bibinfo  {journal}
  {Phys. Rev. Lett.},\ }\textbf {\bibinfo {volume} {105}},\ \bibinfo {pages}
  {040501} (\bibinfo {year} {2010})},\ \Eprint {http://arxiv.org/abs/1003.4662}
  {arXiv:1003.4662} \BibitemShut {NoStop}%
\bibitem [{\citenamefont {Chen}\ \emph
  {et~al.}(2011){\natexlab{b}}\citenamefont {Chen}, \citenamefont {Liu},\ and\
  \citenamefont {Wen}}]{spto_2d}%
  \BibitemOpen
  \bibfield  {author} {\bibinfo {author} {\bibfnamefont {Xie}\ \bibnamefont
  {Chen}}, \bibinfo {author} {\bibfnamefont {Zheng-Xin}\ \bibnamefont {Liu}}, \
  and\ \bibinfo {author} {\bibfnamefont {Xiao-Gang}\ \bibnamefont {Wen}},\
  }\bibfield  {title} {\enquote {\bibinfo {title} {Two-dimensional
  symmetry-protected topological orders and their protected gapless edge
  excitations},}\ }\Doi {10.1103/PhysRevB.84.235141} {\bibfield  {journal}
  {\bibinfo  {journal} {Phys. Rev. B},\ }\textbf {\bibinfo {volume} {84}},\
  \bibinfo {pages} {235141} (\bibinfo {year} {2011}{\natexlab{b}})},\ \Eprint
  {http://arxiv.org/abs/1106.4752} {arXiv:1106.4752} \BibitemShut {NoStop}%
\bibitem [{\citenamefont {Chen}\ \emph
  {et~al.}(2011){\natexlab{c}}\citenamefont {Chen}, \citenamefont {Gu},
  \citenamefont {Liu},\ and\ \citenamefont {Wen}}]{spto_higher}%
  \BibitemOpen
  \bibfield  {author} {\bibinfo {author} {\bibfnamefont {Xie}\ \bibnamefont
  {Chen}}, \bibinfo {author} {\bibfnamefont {Zheng-Cheng}\ \bibnamefont {Gu}},
  \bibinfo {author} {\bibfnamefont {Zheng-Xin}\ \bibnamefont {Liu}}, \ and\
  \bibinfo {author} {\bibfnamefont {Xiao-Gang}\ \bibnamefont {Wen}},\
  }\href@noop {} {\enquote {\bibinfo {title} {Symmetry protected topological
  orders and the cohomology class of their symmetry group},}\ } (\bibinfo
  {year} {2011}{\natexlab{c}}),\ \Eprint {http://arxiv.org/abs/1106.4772}
  {arXiv:1106.4772} \BibitemShut {NoStop}%
\end{thebibliography}
%

\end{document}